\documentclass[graybox]{svmult}

\usepackage{mathptmx}       % selects Times Roman as basic font
\usepackage{helvet}         % selects Helvetica as sans-serif font
\usepackage{courier}        % selects Courier as typewriter font
\usepackage{type1cm}        % activate if the above 3 fonts are
\usepackage{amssymb}       % selects Times Roman as basic font
\usepackage{makeidx}         % allows index generation
\usepackage{graphicx}        % standard LaTeX graphics tool
\usepackage{multicol}        % used for the two-column index
\usepackage[bottom]{footmisc}% places footnotes at page bottom

\begin{document}
\title*{Phase-Field Modeling of Nonlinear Material Behavior}
\author{Y.-P.\ Pellegrini, C.\ Denoual and L.\ Truskinovsky}
\institute{Y.-P.\ Pellegrini and C.\ Denoual \at CEA, DAM, DIF,
F-91297 Arpajon, France; \email{yves-patrick.pellegrini@cea.fr,\newline
christophe.denoual@cea.fr}\and L.\ Truskinovsky\at Laboratoire de
M\'ecanique des Solides, CNRS UMR-7649, \'Ecole Polytechnique, Route
de Saclay, F-91128 Palaiseau Cedex, France; 
\email{trusk@lms.polytechnique.fr}\and
{\bf In:} K.\ Hackl (ed.) \emph{Proceedings of the IUTAM Symposium on Variational Concepts with Applications to the Mechanics of Matrerials, September 22--26, 2008, Bochum, Germany} (Springer-Verlag, presumably 2010). In press.}

\maketitle

\abstract{Materials that undergo internal transformations are usually described in
solid mechanics by multi-well energy functions that account for both
elastic and transformational behavior. In order to separate the two effects,
physicists use instead phase-field-type theories where conventional linear elastic
strain is quadratically coupled to an additional field that describes the evolution
of the reference state and solely accounts for nonlinearity. In this paper we
propose a systematic method allowing one to split the nonconvex energy into harmonic
and nonharmonic parts and to convert a nonconvex mechanical problem into a partially
linearized phase-field problem. The main ideas are illustrated using the simplest
framework of the Peierls--Nabarro dislocation model.}

%%%%%%%%%%%%%%%%%%%%%%%%%%%%%%%%%%%%%%%%%%%%%%%%%%%%%%%%%%%%%%%%
%%%%%%%%%%%%%%%%%%%%%%%%%%%%%%%%%%%%%%%%%%%%%%%%%%%%%%%%%%%%%%%%

\section{Introduction}
\label{sec:1} Nonconvex energy potentials are used in solid
mechanics for the modeling of martensitic transformations
\cite{E75}, plasticity \cite{CARP05} and fracture \cite{TRUS96}.
Parts of the resulting energy landscapes correspond to sufficiently
smooth deformations preserving the locally affine structure of the
lattice environment of each atom. Other parts represent highly
distorted atomic arrangements associated with either loss or
reacquisition of nearest neighbors. While deformations of the first
type can (often) be described by the conventional strain tensor of
(linear) elasticity theory, a representation of the deformations of
the second type requires introducing additional internal
variables accounting for deviations from the local affinity of the
stressed atomic configurations. In particular, these supplementary
variables describe the evolution of the local reference state (LRS)
from which the elastic deformations are measured
\cite{CHOK99,HAKI05,WANG97}. The main difference between the elastic strains
and these supplementary internal variables is that the dynamics of
the former is typically inertial, while that of the latter is usually overdamped.
Sometimes the nonelastic variables can be minimized out as in the case of
deformational plasticity (e.g., \cite{CARS02}).
In this paper we deal instead with situations where the internal
variables have to revealed rather than hidden.

We assume that the coarse-grained nonconvex energy density $f(\varepsilon)$ is
known either from extrapolations of experimental measurements or from
\emph{ab-initio} calculations involving atomic homogeneity constraints.
We suppose that the argument $\varepsilon$ of this function, that represents
a coarse-grained strain, is small and can be additively split into the
linear elastic part $e$, and a phase-field part $\eta$ that accounts
for the nonelastic evolution of the LRS. Our next assumption is
that $f$ can be represented as a sum of two terms:
the elastic energy $f_e$, which depends on $e=\varepsilon-\eta$ and  the phase-field
energy $g$, which depends on $\eta$. We interpret $f(\varepsilon)$ as
the outcome of adiabatic elimination of the variable $\eta$ and consider the inverse
problem of recovering the phase-field energy $g(\eta)$ from the function
$f(\varepsilon)$ under the assumption that the function $f_e(e)$ is quadratic.
The problem of the identification of $g(\eta)$ reduces to a problem of
optimization and the relation between the 'optimally'
related functions $f(\varepsilon)$ and $g(\eta)$ is studied in some prototypical
cases. If, in contrast, the function  $g(\eta)$ is chosen independently, the corresponding
function $f(\varepsilon)$ is typically non-smooth and non single-valued, e.g.
\cite{DELP01}.

To motivate the need for the phase-field variables we consider in
full detail a specific physical example. It deals with the mixed,
discrete-continuum representation of a dislocation core
\cite{HIRT82,ORTI99}. More specifically, we develop a modified
version of the classical Peierls--Nabarro (PN) model that accounts
for a finite thickness of the slip region. In this problem the
coarse-grained description of the slip zone is provided by the
so-called $\gamma$-potential \cite{CHRI70,WOOD05}. The phase field
represents an ``atomically sharp'' slip and the part of the
interaction potential related to $g$ gives rise to the slip-related
pull-back force \cite{DENO04,ORTI99,SUN93}. Our general method of
recovering the expression for this force represents an extension of
Rice's transform, which was first introduced in the context of a
dislocation nucleation problem \cite{RICE92}.

In this paper only the simplest scalar problem in a one-dimensional setting is
considered. The slightly more general question of extracting
from the coarse-grained energy a convex (instead of quadratic)
component will be examined elsewhere \cite{PELL09}.

\section{Surface Problem}
\label{sec:surfprobl}
We begin with the special case when the phase field is
localized on a surface. In problems involving fracture or slip it often proves
convenient to represent the energy of a body as the sum of a bulk term depending on
strain gradients and a surface term penalizing displacement discontinuities. The bulk
term is usually modeled by linear elasticity. The modeling of the surface energy is
less straightforward \cite{DELP01,TRUS96}. For instance, the models will be different depending
on whether the location of the discontinuities is known a priori or not.

In a 1D setting with a \emph{known} fracture set the equilibrium problem reduces to
minimizing the following energy functional
\begin{equation}
\label{11} W[u]=\int_0^1 \mathrm{d}x\, f_e(u_x)+\sum_{\Gamma_a}
f_a\bigl(\delta(x)\bigr).
\end{equation}
Here $f_e(e)=(E/2)e^2$ , $E>0$ is the elastic modulus and  $a$ is a
coarse-graining length scale that typically exceeds several atomic sizes. The set
$\Gamma_a$ in (\ref{11}) represents discontinuity points
resolved at scale $a$ and
 $\delta(x)=[u]_a (x)$ is the corresponding displacement discontinuity.
 The surface energy $f_a(\delta)$ is then an effective interaction over the
distance $a$; in particular, the shear-related component of $f_a(\delta)$
coincides the $\gamma$-potential mentioned in the Introduction.

In the case when the fracture set is \emph{unknown} the surface energy has to be chosen differently.
The reason is that in this model the displacement discontinuity at scale $a$ does not represent the
microscopic slip between neighboring atomic planes, and therefore the difference between
elastic deformation and inelastic slip has to be yet resolved at this scale \cite{RICE92}. More precisely,
linear elasticity, which has nothing to do with slip and which is already accounted for
in the bulk term, has not been excluded from $f_a(\delta)$. The
identification  of the surface energy in (\ref{11}) with $f_a(\delta)$, which is
quadratic at the origin, leads in a free discontinuity problem to a degenerate solution with
infinitely many infinitely small discontinuities \cite{DELP01}.

To remove linear elasticity from the surface term, one should replace the
coarse-grained discontinuity $[u]_a$ by the atomically sharp slip
$\eta(x)=[u](x)$ that does not depend on $a$. The energy (\ref{11})
is then rewritten as
\begin{equation}
W[u]=\int_0^1 \mathrm{d}x\, f_e(u_x)+\sum_\Gamma g(\eta),
\end{equation}
where now $\Gamma$ is the set of discontinuity points corresponding to $a=0$. The
problem is to find the relation between the function $f_a(\delta)$, representing
an empirical input, and the unknown function $g(\eta)$.

To define $g(\eta)$ we divide the total slip $\delta$
into an elastic part, $a\, e$, where $e$ is an equivalent elastic strain, and an
inelastic part $\eta$. The function $g(\eta)$ is defined by the condition that
$f_a(\delta)$ is a relaxation of the energy $a f_e(e)+g(\eta)$ under the condition
that $a e+\eta=\delta$, namely:
\begin{equation}
\label{eq:defg}
f_a(\delta)=\inf_{\eta}
\left[a\frac{E}{2}\left(\frac{\delta-\eta}{a}\right)^2+g(\eta)\right].
\end{equation}
If the energy $f_a(\delta)$ is a single-well function and the infimum is unique, the function $g$ is completely defined. If $f_a(\delta)$ is periodic as in the case of dislocations, in order to have a uniquely-defined $g(\eta)$, we need to replace in definition (\ref{eq:defg}) the global minimization by a properly-defined local minimization denoted hereafter by '$\mathop{\mathrm{inf}}_{\mathrm{loc}}$'(minimization over $\eta$ starting from the minimum of $f_a$ closest to $\delta$.)

In what follows, our task will be to reverse definition (\ref{eq:defg}) and to recover the nonequilibrium energy
$g(\eta)$ from $f_a(\delta)$. What allows us to proceed is the specific (harmonic)
structure of the elastic part of the energy.

We observe that the function $g$ must satisfy the following necessary condition
\begin{equation}
(E/a)(\delta-\eta)=g'(\eta).
\end{equation}
Moreover, differentiation of (\ref{eq:defg}) wrt.\ $\delta$ gives
\begin{equation}
f'_a(\delta)=(E/a)(\delta-\eta).
\end{equation}
These two equations allow one to represent $g'(\eta)$ in the following
parametric form \cite{DENO04,DENO07,RICE92}
\begin{equation}
\label{eq:paramgprime}
\bigl(\eta,g'(\eta)\bigr)=\left(\delta-(a/E)f'_a(\delta),f'_a(\delta)\right).
\end{equation}
The parametric representation for $g(\eta)$ then reads
\begin{equation}
\label{eq:paramg} \bigl(\eta,g(\eta)\bigr)=\left(\delta-(a/E)\,f'_a(\delta),
f_a(\delta)-[a/(2E)]\left[f'_a(\delta)\right]^2\right).
\end{equation}
Since for  nonconvex $f_a(\delta)$ this representation may lead to a multivalued
function $g(\eta)$ formula (\ref{eq:paramg}) must be supplemented by an additional
branch selection procedure.
%--------------------------------------------------------------
\begin{figure}[ht]
\begin{center}
\includegraphics[height=3.9cm]{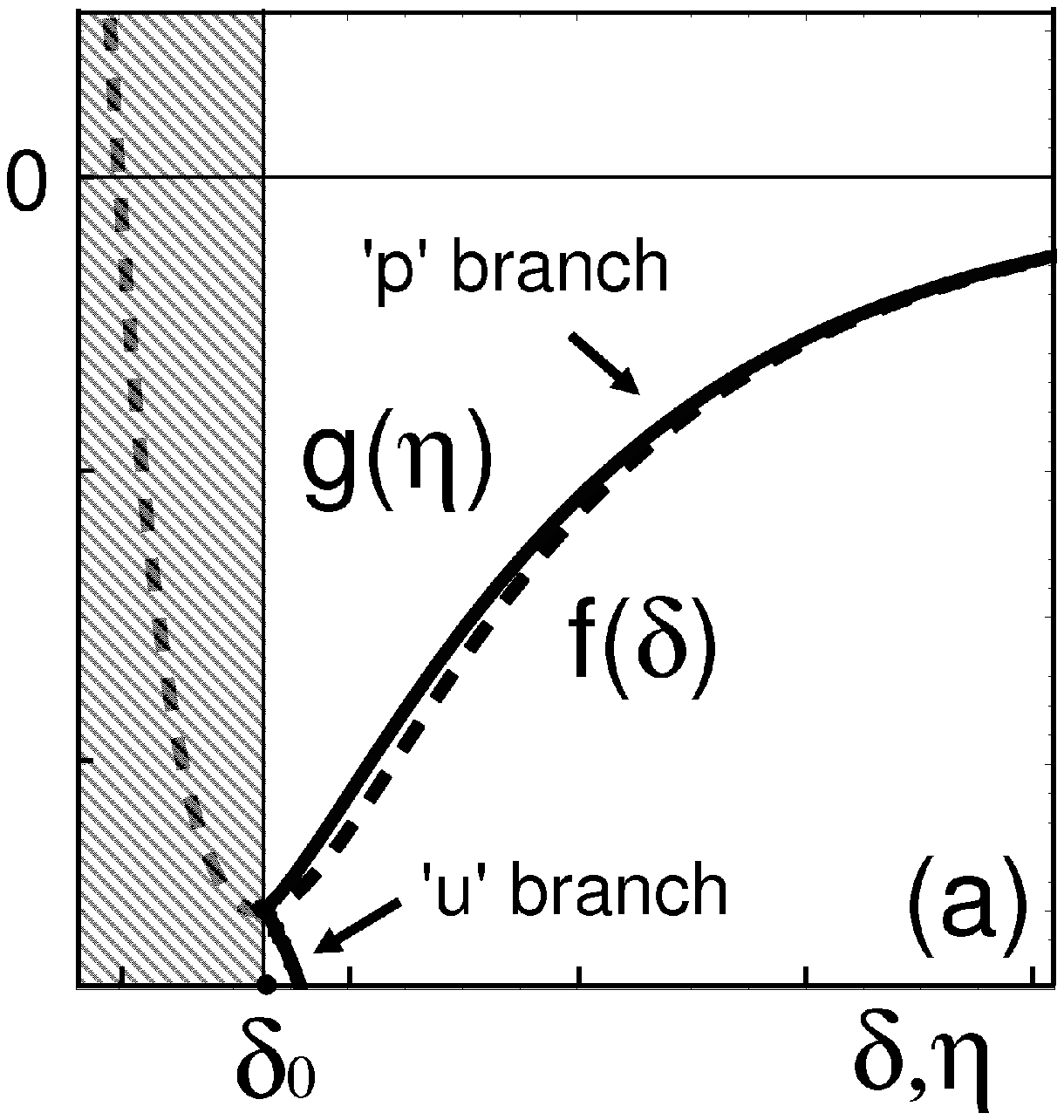}\,
\includegraphics[height=3.9cm]{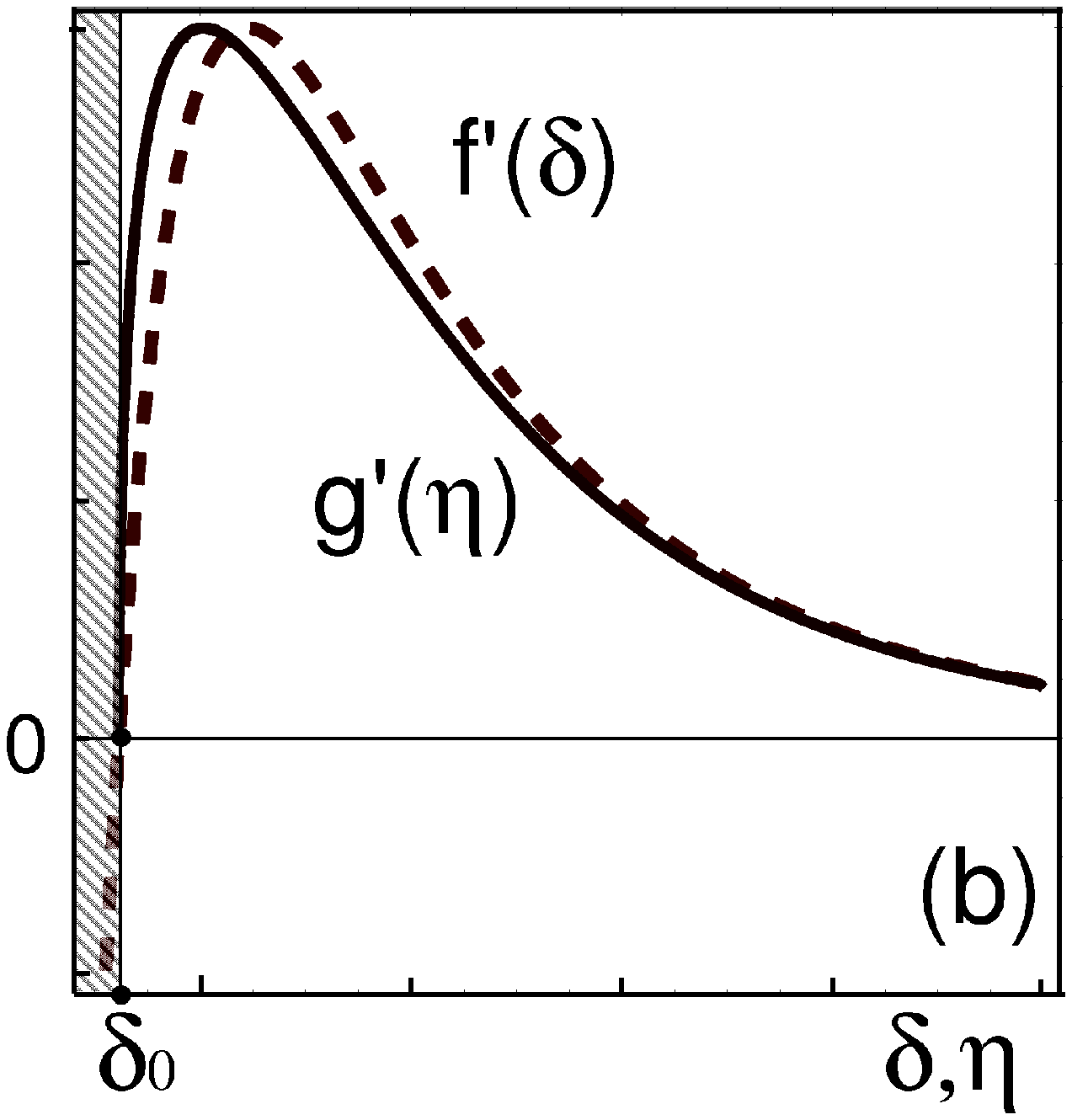}\,
\includegraphics[height=3.9cm]{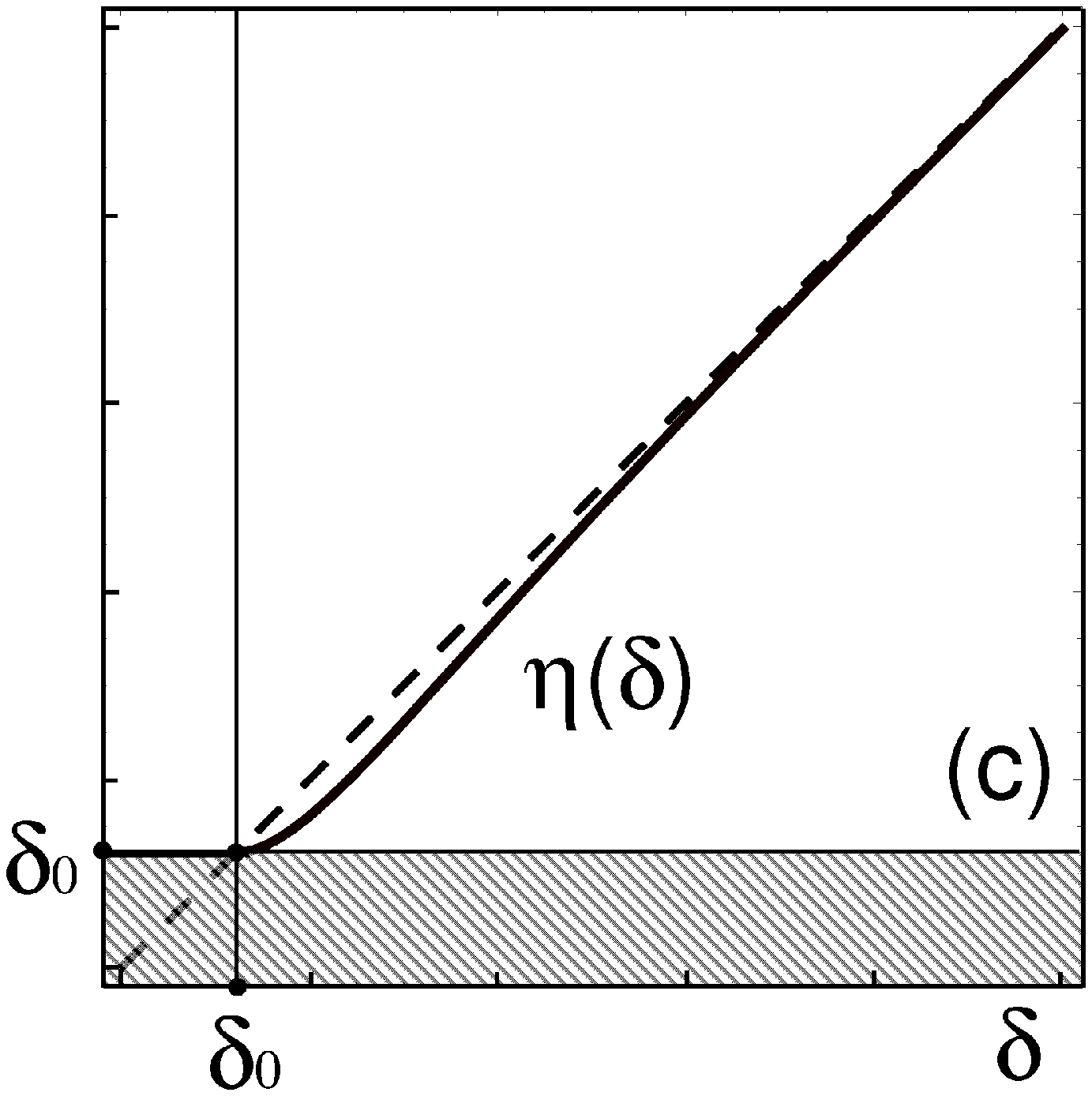}\
\caption{\label{fig:lenj}Parametric transforms
(\ref{eq:paramgprime}), (\ref{eq:paramg}) applied to a Lennard-Jones
potential $f(\delta)=\delta^{-12}-\delta^{-6}$. a) $f(\delta)$ (dashed), and  the two branches
of $g(\eta)$ (solid), where `p' (`u') labels the physical
(unphysical) branch; b) $f'(\delta)$ and the `p' branch of
$g'(\eta)$; c) $\eta(\delta)$.}
\end{center}
\end{figure}
%--------------------------------------------------------------
To illustrate the mapping $f(\delta) \rightarrow g(\eta)$ given by (\ref{eq:paramg})
and the selection of a physical branch we consider a Lennard-Jones potential $f_a$,
with $a=1$ and assume that $E=f''(\delta_0)$, where $\delta_0$ is the only minimum of
$f$ (see Fig. \ref{fig:lenj}). Notice that the resulting function $g'(\eta)$ has an
infinite slope at $\eta=\delta_0$ and that for $\eta\gtrsim\delta_0$ we must have
$g(\eta)\propto (\eta-\delta_0)^{3/2}$.

The removal of the linear elastic part of the energy becomes
important in PN-type modeling of dislocations. Consider,
for instance, a straight screw dislocation in an isotropic
linear-elastic body and assume that the sharp discontinuity plane,
$y=0$, lies between the two effective gliding surfaces located at
$y=\pm a/2$. To account for the finite thickness of the core region
$a$ we need to modify the classical PN model \cite{HIRT82}.
According to our interpretation the linear elastic stress outside the slip
region $(-a/2,a/2)$ must be balanced by the coarse-grained pull-back
stress that is resolved at the spatial scale $a$. We therefore
interpret the pull-back stress at this scale as $f'_a\bigl(\delta(x)\bigr)$, where
$f_a$ is the $\gamma$-potential, a periodic function with period $b$
and with $f'_a(0)=0$. The expression for the linear stress outside the slip region is
derived in the Appendix. With these considerations in mind we obtain for the unknown function $\eta(x)$ representing a
mathematical slip at $y=0$ the following system of equations
\begin{eqnarray}
\label{eq:nlpn1} &&\hspace{-1.5em}-\frac{\mu}{\pi
a}\int_{-\infty}^{+\infty} {\rm
d}x'\,\eta'(x')\arctan\frac{a}{2(x-x')}+\overline{\sigma}_a(x)=f'_a\bigl(\delta(x)\bigr),\\
\label{eq:nlpn2} &&\hspace{7em}\delta(x)=(a/\mu)
f'_a\bigl(\delta(x)\bigr)+\eta(x),
\end{eqnarray}
where $\overline{\sigma}_a$ is the resolved applied stress at scale $a$. If we match
the linear elastic behavior at $\eta=0$ with that in the bulk regions we obtain that
$\mu=a f''_a(0)$. Using in this relation the physical shear modulus and the value of
$f''_a(0)$ from the $\gamma$-potential provides a rough estimate for $a$, the
effective interaction range.

We notice that parameter $a$ enters both equations
(\ref{eq:nlpn1},\ref{eq:nlpn2}), which makes this system different from the one
studied in \cite{ORTI99,RICE92}. The ideas behind our nonlocal extension of the PN
model are also different from that of Ref.\ \cite{MILL98} where a nonlocal kernel
was introduced empirically as part of the pull-back stress, and the usual
$1/(x-x')$ kernel was used for the bulk stress.

To bring the system (\ref{eq:nlpn1},\ref{eq:nlpn2}) into the framework of
phase-field models, we identify the effective pull-back force
$f'_a\bigr(\delta(\eta)\bigl)$ with $g'(\eta)$ and rewrite Eq.\ (\ref{eq:nlpn1}) as
\begin{equation}
\label{eq:nlpnad}\frac{\mu}{2}\int_{-\infty}^{+\infty}\hspace{-1em}
{\rm
d}x'\,K_a(x-x')\,\eta'(x')+\overline{\sigma}_a(x)=g'\bigl(\eta\bigr).
\end{equation}
where $K_a(x)=-(2/\pi a)\arctan (a/2 x)$. It is now easy to see that $g'(\eta)$
enjoys the parametric representation
\begin{equation}
\label{eq:paramgprime2}\bigl(\eta,g'(\eta)\bigr)
=\left(\delta-\frac{a}{\mu}f'_a(\delta),f'_a(\delta)\right),
\end{equation}
where we recognize the mapping (\ref{eq:paramgprime}) (see also \cite{ORTI99,RICE92,SUN93}).
To make the link with the classical PN model one needs to consider the limit $a\to
0$. By computing $\eta$ in terms of $\delta$ and  expanding (\ref{eq:nlpnad}) in
powers of $a$, we obtain to order $O(a)$ the following `gradient' extension of the PN model
\begin{equation}
\label{eq:nlpngrad} -\frac{\mu}{2\pi}\int_{-\infty}^{+\infty} {\rm
d}x'\,\frac{\delta'(x')}{x-x'}+\lambda\delta''(x)+
\widetilde{\sigma}_a(x)=f\bigl(\delta(x)\bigr),
\end{equation}
where $\lambda= a\mu/4$. For different weakly or strongly nonlocal generalizations of the PN model see
\cite{MILL98,ROSA01}. Equation (\ref{eq:nlpngrad}) features an effective applied
stress that differs from $\sigma_a(x,0)$ defined in the Appendix by an $O(a)$
correction, namely $\widetilde{\sigma}_a(x)\equiv
\sigma_a(x,0)+(a/2)\left[(1/2)\partial_y \sigma_a(x,0)-K_0\star\partial_x
\sigma_a(x,0)\right]$. The classical PN model is retrieved by letting $a=0$.

%%%%%%%%%%%%%%%%%%%%%%%%%%%%%%%%%%%%%%%%%%%%%%%%%%%%%%%%%%%%%%%%
\section{Bulk Problem}
\label{subsec:bulk} Now let us place the problem in a more general
framework. The task is to approximate locally the empirical
potential $f(\varepsilon)$ by a quadratic function with an optimally
chosen reference state $\eta$, and to associate with this state a reference energy $g(\eta)$.
Behind such construction is the assumption that all the
nonlinearity of the problem is related to the evolution of the
reference state. The simplest setting to pose formally the problem
is the one-dimensional geometrically linearized theory of nonlinear elastic bars.

According to our interpretation the empirical energy is represented as
\begin{equation}
\label{eq:relax} f(\varepsilon)=\inf_{\eta,{\rm
loc}}\left[\frac{E}{2}(\varepsilon-\eta)^2+g(\eta)\right]
\end{equation}
and the problem is to find the intrinsic phase-field function $g(\eta)$. Following
the previous section we write the parametric representation for $g'(\eta)$ in the
form
\begin{equation}
\label{eq:paramgprime3} \bigl(\eta,g'(\eta)\bigr)
=\left(\varepsilon-\frac{f'(\varepsilon)}{E},f'(\varepsilon)\right).
\end{equation}
The function $g(\eta)$ is then given by the mapping
\begin{equation}
\label{eq:paramg3} \bigl(\eta,g(\eta)\bigr)
=\left(\varepsilon-\frac{f'(\varepsilon)}{E},f(\varepsilon)-\frac{f'(\varepsilon)^2}{2E}
\right).
\end{equation}
The consistency of this procedure requires the parameter $E$ and the function
$f(\varepsilon)$ to be related. If we expand the parametric definition of $g(\eta)$
near a reference state $\varepsilon_0$ where $f'(\varepsilon_0)=0$, we obtain
$g(\varepsilon_0)=f(\varepsilon_0)$, $g'(\varepsilon_0)=0$ and
$g''(\varepsilon_0)=f''(\varepsilon_0)/[1-f''(\varepsilon_0)/E]$. The natural choice
$E=f''(\varepsilon_0)$ makes $g''(\varepsilon_0)$ infinite. The behavior of the
higher derivatives of $g(\eta)$ near $\eta=\varepsilon_0$ is found by assuming
(without loss of generality) that derivatives $f^{(k)}(\varepsilon_0)$ vanish for
$k=3,\ldots,n-1$. The order of the asymptotics depends on $n>2$, which is the first
integer such that $f^{(n)}(\varepsilon_0)\not=0$:
\begin{equation}
\label{eq:etagexpan}
\bigl(\eta,g(\eta)\bigr)\simeq\left(\varepsilon_0-\frac{f^{(n)}(\varepsilon_0)}{(n-1)!
E}\delta\varepsilon^{(n-1)},
f(\varepsilon_0)-\frac{(n-1)f^{(n)}(\varepsilon_0)}{n!}\delta\varepsilon^n\right).
\end{equation}
Hence $g$ behaves near its minimum as:
$|g(\eta)-g(\varepsilon_0)|\sim|\eta-\varepsilon_0|^{n/(n-1)}$. The generic
case is $n=3$; the case $n=4$ corresponds to periodic potential relevant for
dislocations; for $f$ locally harmonic, $n=\infty$.
%--------------------------------------------------------------
\begin{figure}[ht]
\begin{center}
\includegraphics[height=5cm]{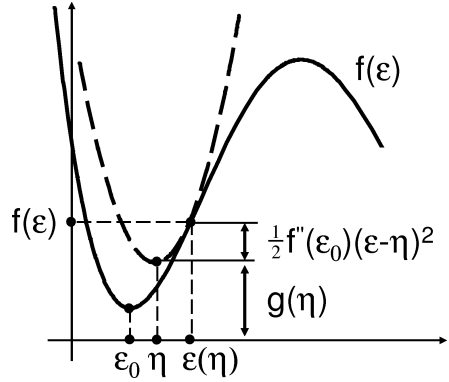}
\caption{\label{fig:transform}Geometrical illustration of the construction defined by
Eqs.\ (\ref{eq:paramgprime3}-\ref{eq:fstar}). }
\end{center}
\end{figure}
%--------------------------------------------------------------

Observe now that the function $g$ computed from
(\ref{eq:paramgprime3},\ref{eq:paramg3}), can also be viewed as a solution of the
following optimization problem:
\begin{equation}
\label{eq:fstar} g(\eta)=\sup_{\varepsilon, {\rm
loc}}\left[f(\varepsilon)-\frac{E}{2}(\varepsilon-\eta)^2\right],
\end{equation}
which is a natural inverse of (\ref{eq:relax}) (see also \cite{PONT02,ROCK97}). Since
the equation $\eta=\varepsilon-f'(\varepsilon)/E$ may have several solutions
$\varepsilon(\eta)$, the representation (\ref{eq:fstar}) removes the ambiguity by
always selecting the upper branch.
%--------------------------------------------------------------
\begin{figure}[ht]
\begin{center}
\includegraphics[height=4.1cm]{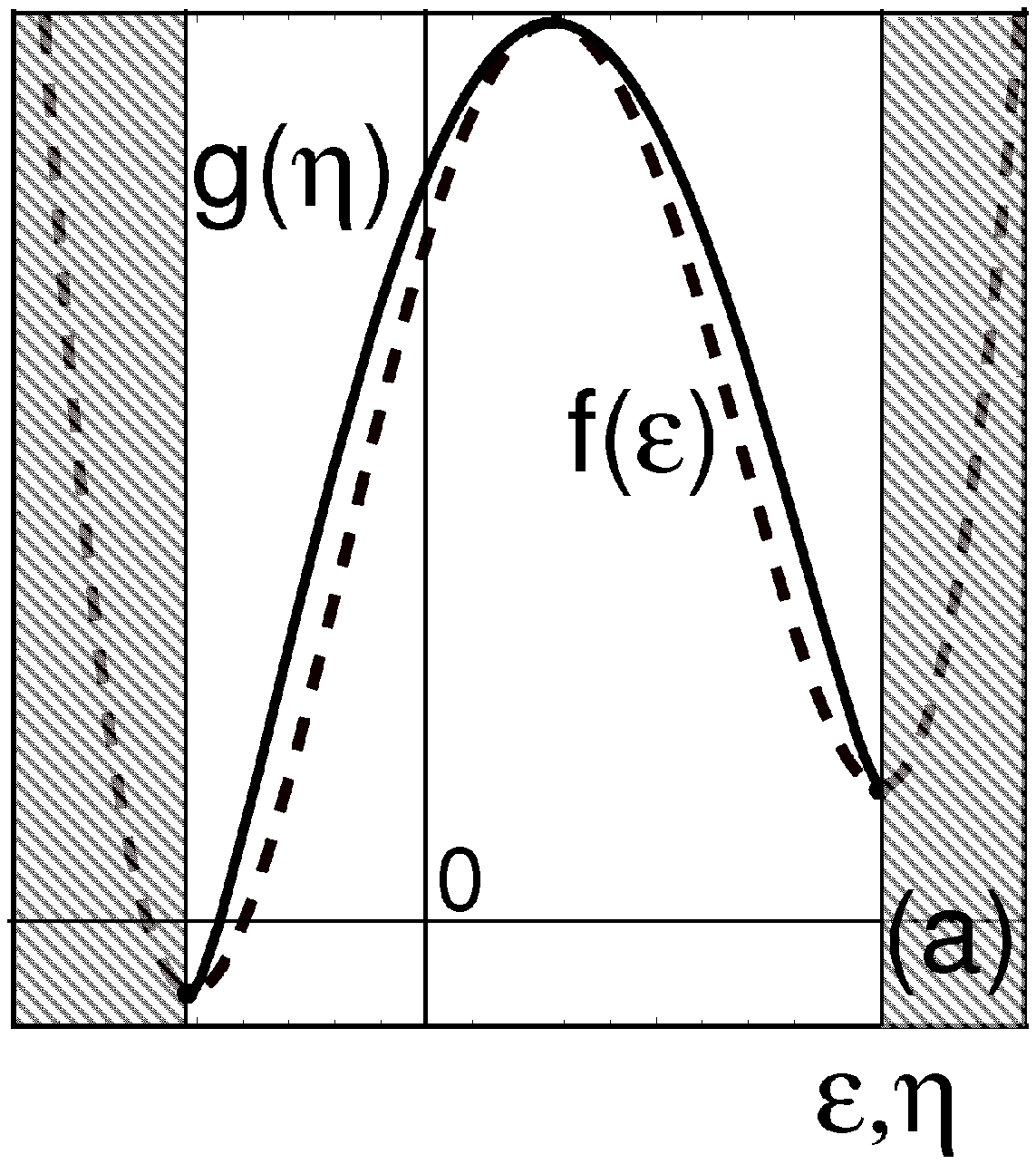}\quad
\includegraphics[height=4.1cm]{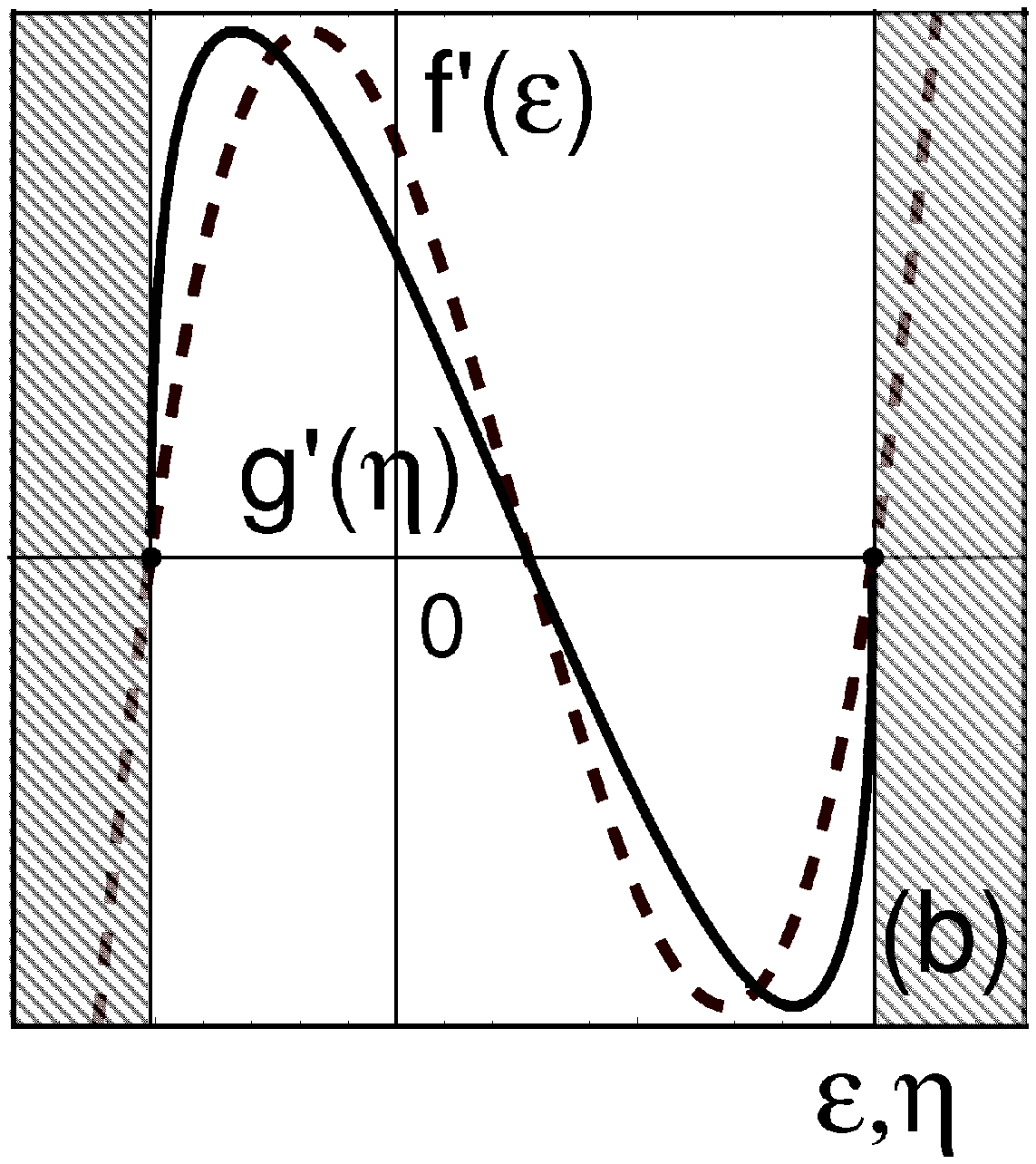}\quad
\includegraphics[height=4.1cm]{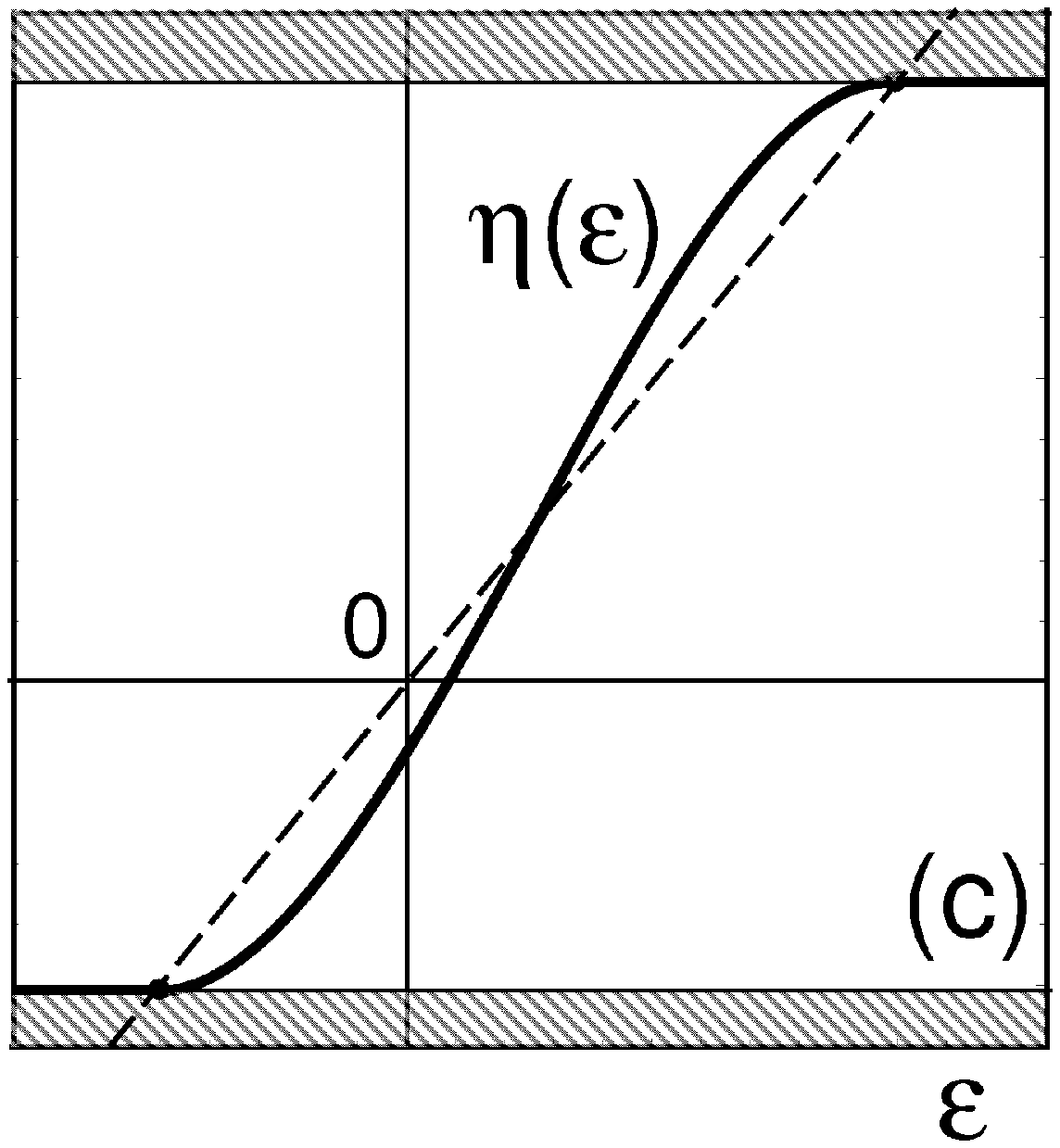}\
\caption{\label{fig:doublewell1d} Phase-field representation of a  double-well potential
$f(\varepsilon)=(\varepsilon-1)^2(2\varepsilon+1)^2+0.2
\varepsilon$.}
\end{center}
\end{figure}
%--------------------------------------------------------------

The working of Eqs.\ (\ref{eq:paramgprime3}-\ref{eq:fstar}) with
$E=f''(\varepsilon_0)$ is illustrated in Fig.\ \ref{fig:transform}. In the domain at
the left of $\varepsilon_0$, where $f$ grows faster than harmonic the desired
tangency point does not exist. In this case the difference
$f(\varepsilon)-\frac{E}{2}(\varepsilon-\eta)^2$ is maximized at
$\varepsilon=-\infty$. This situation takes place in the Lennard-Jones example of
Sec.\ \ref{sec:surfprobl} where we have to use $g(\delta)=+\infty$ for
$\delta<\delta_0$ (hatched area of Fig.\ \ref{fig:lenj}a).

To handle general multi-well energies, we first introduce the Stillinger-Weber mapping
$\varepsilon_0(\varepsilon)$ that links to any state $\varepsilon$ the local minimum
$\varepsilon_0$ of $f(\varepsilon)$ that would be attained from this state by
steepest-descent \cite{STIL84}. Next, we modify equations (\ref{eq:relax}) and
(\ref{eq:fstar}) as:
\begin{eqnarray}
\label{eq:relax2} f(\varepsilon)=\inf_{\eta,{\rm loc}}\left[\frac{1}{2}
f''\bigl(\varepsilon_0(\varepsilon)\bigr)(\varepsilon-\eta)^2+g(\eta)\right],\\
\label{eq:fstar2} g(\eta)=\sup_{\varepsilon, {\rm
loc}}\left[f(\varepsilon)-\frac{1}{2}f''\bigl(\varepsilon_0(\eta)\bigr)(\varepsilon-\eta)^2\right].
\end{eqnarray}
Whereas (\ref{eq:fstar2}) defines $g$, equation (\ref{eq:relax2})
states that knowing $f$ is equivalent to knowing $g$ plus
the linear-elastic behavior of $f$ near its local minima.

The precise meaning of the ``loc''  in Eqs.\ (\ref{eq:relax2}),
(\ref{eq:fstar2}) is as follows. Operationally, the minimization in the definition
of $f$ is carried out over $\eta$, starting from $\varepsilon_0(\varepsilon)$, the local
minimum nearest to $\varepsilon$ determined by the SW mapping; the corresponding
elastic modulus is also determined by the starting point. The maximization in the
definition of $g(\eta)$ proceeds along similar lines except that now the relevant
elastic modulus is determined by the local minimum closest to $\eta$, and is fixed
during the maximization. The optimization is carried out starting from
$\varepsilon=\eta$.

Figs.\ \ref{fig:doublewell1d} illustrate the case of a double-well potential with
unequal curvatures of the wells. Notice that in contrast to what we saw in Fig.\
\ref{fig:lenj} the function $\eta(\varepsilon)$ is now bounded.

%--------------------------------------------------------------
\begin{figure}[ht]
\begin{center}
\includegraphics[height=4.1cm]{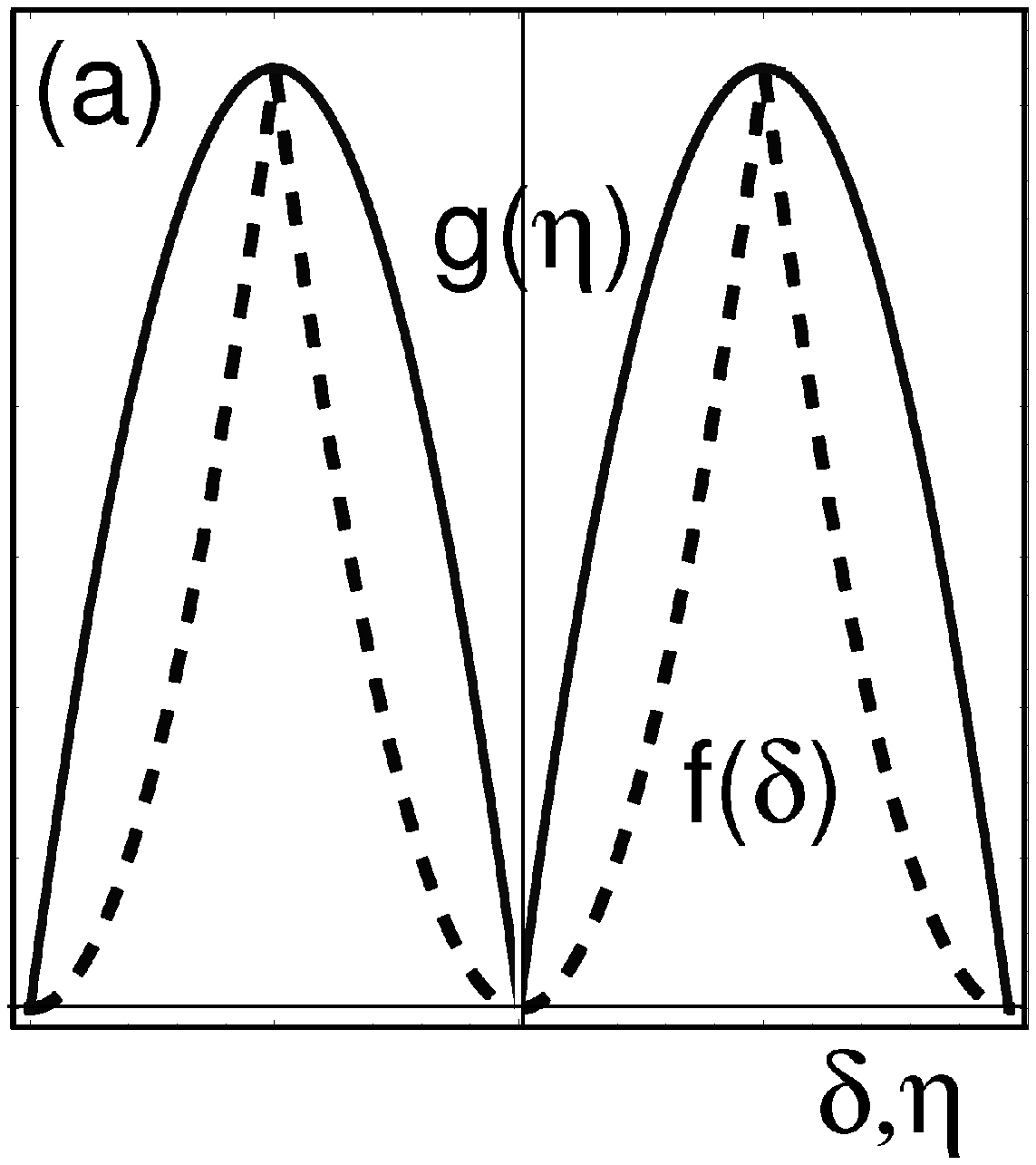}\quad
\includegraphics[height=4.1cm]{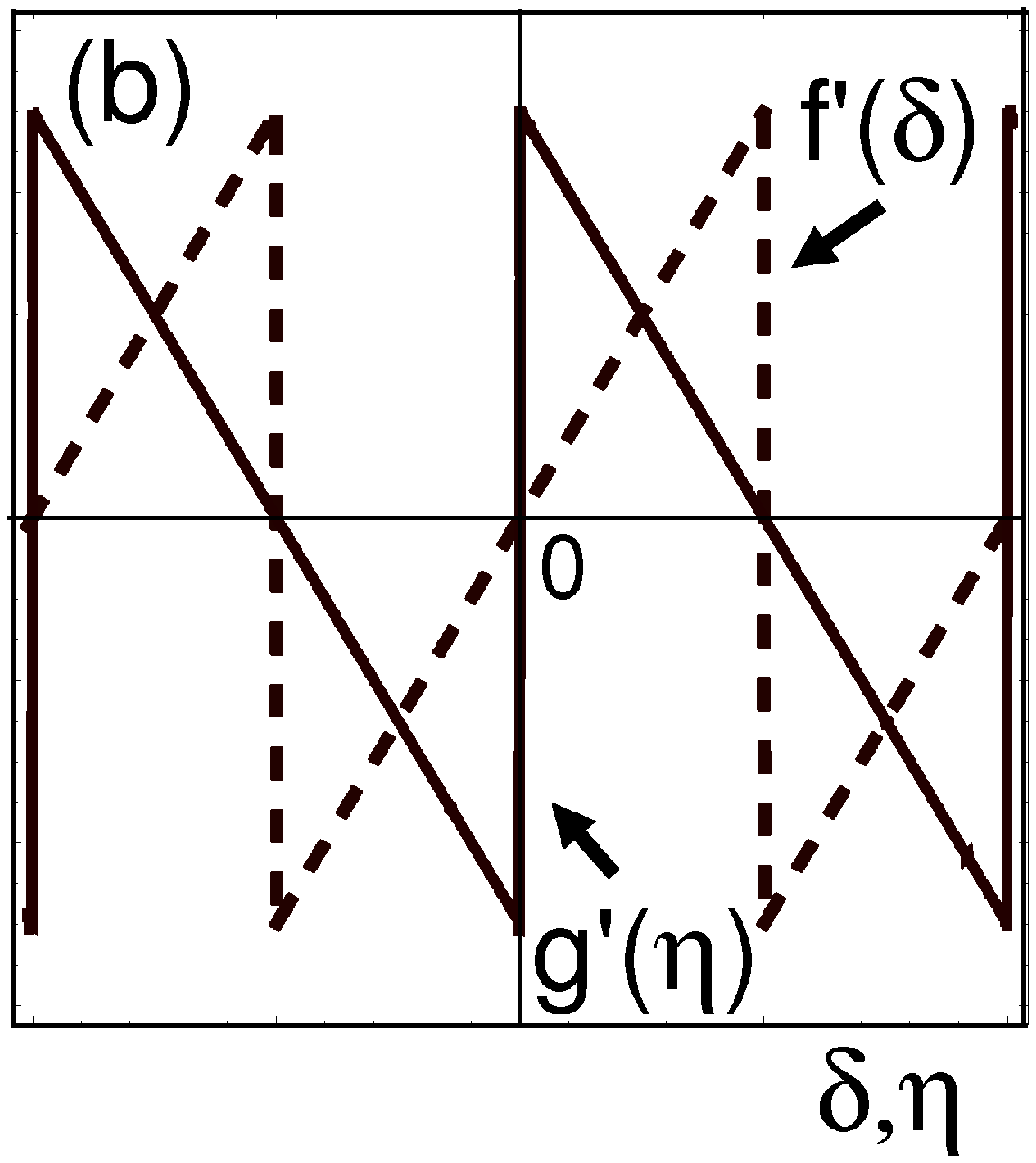}\quad
\includegraphics[height=4.1cm]{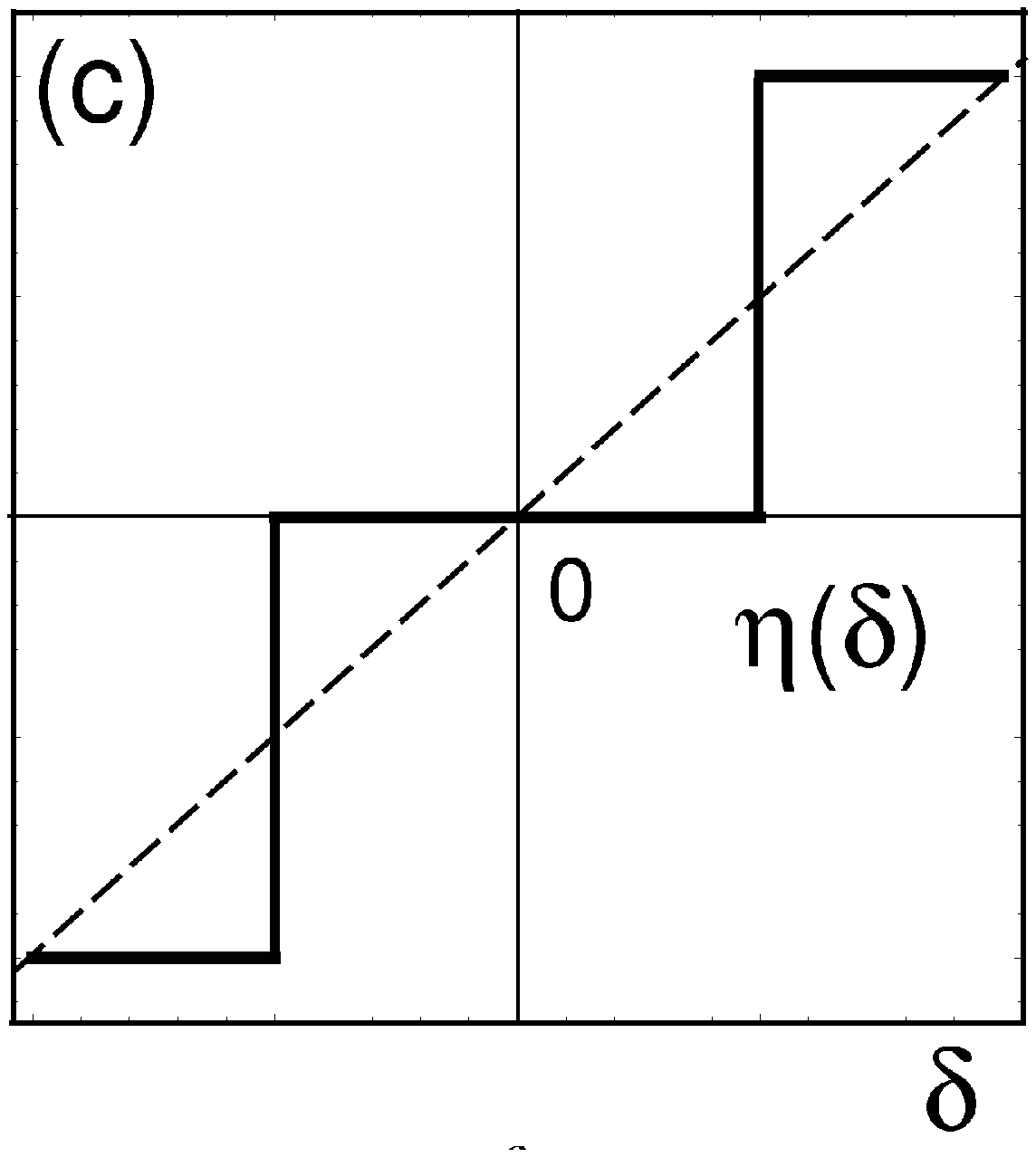}\
\caption{\label{fig:php} Transform (\ref{eq:fstar2}) applied to the
piecewise-harmonic periodic potential
$f(\delta)=(\delta-[\delta])^2/2$. Notation $[\cdot]$ stands for the integer
part. The function $\eta(\delta)$ is determined as the argmin in
(\ref{eq:relax2}).}
\end{center}
\end{figure}
%--------------------------------------------------------------

Another interesting case is the periodic potential that is used in the
description of reconstructive phase transitions (e.g., \cite{CZ04}).
Consider, for instance, the piecewise-harmonic \emph{periodic}
case shown in Fig.\ \ref{fig:php} that is often used in analytical
studies \cite{MILL98,ROSA01}. The parametric representation
(\ref{eq:paramgprime3}) of $g$ is here useless and the definition
(\ref{eq:fstar2}) must be used instead. In this extreme case,
\emph{all} elasticity has been removed from $g$ and the resulting
$g(\eta)$ is cone-shaped at its minima (Fig.\ \ref{fig:php}a)
as predicted by Eq.\ (\ref{eq:etagexpan}) for $n\to\infty$.
The force $g(\eta)$ is discontinuous (Fig.\ \ref{fig:php}b) and its
extreme values provide thresholds for the evolution of $\eta$, whose
stepwise character is an artifact due to the absence of smooth
spinodal regions in $f$.

It is also instructive to consider for
comparison the case of an unbounded harmonic potential
$f(\varepsilon)=(E_f/2)(\varepsilon-\varepsilon_0)^2$. From
(\ref{eq:fstar}) with $E=E_f$, one deduces that $g(\eta)=+\infty$ if
$\eta\not=\varepsilon_0$ and $g(\eta=\varepsilon_0)=0$. This trivial
 example indicates that in a purely linear-elastic model, the reference state does not have to evolve.

\section{Concluding Remarks}
The goal of this paper was to reveal in the simplest 
setting the variational nature of the generalized Rice transform. The problem consists in splitting 
a coarse-grained lattice potential $f$,  describing the overall deformation of a sufficiently large number of atoms,  into a (quasi) convex elastic potential and an inelastic potential $g$ dealing with structural rearrangements. Here the potential $f$ is assumed to be measurable by molecular
statics along a prescribed deformation path relevant to the material transformation in question. For simplicity, the elastic potential is assumed in this paper to be a standard quadratic function of the macroscopic strain. The inelastic potential $g$  must be a function of the phase-field variable $\eta$, whose identification represents an important part of the problem.
While our precise construction solving the above problem is presented in the static setting (see Eq.\ (\ref{eq:fstar2})), the motivation for the splitting concerns, first of all, dynamical applications (e.g., \cite{DENO04}). Thus we assume that material displacement $u$ associated to the strain $\varepsilon=\partial_x u$ evolves inertially almost without damping (standard elastodynamics), while the dynamics of the phase-field variable is overdamped. More precisely we assume that the relaxation of the variable
 $\eta$ follows the time-dependent Ginzburg-Landau (TDGL) equation. By means of an empirical `viscosity' parameter $\nu$ we can write the evolution equation in the form
$$
\dot{\eta}=-\frac{1}{\nu}\frac{\partial}{\partial\eta}\left[\frac{1}{2}
f''\bigl(\varepsilon_0(\varepsilon)\bigr)(\varepsilon-\eta)^2+g(\eta)\right],
$$
where we have omitted for simplicity the conventional gradient-penalizing terms  (e.g., \cite{DENO04,TRUS93}). 
In the static setting the above equation reduces to our basic Equ.\ (\ref{eq:relax2}).
The definitive advantage of separating the wave motion from an
overdamped TDGL relaxation is the possibility to attribute effective damping only to large atomic displacements. 
Our preliminary 
investigations \cite{PELL09} indicate that extending the variational set-up presented in this paper to higher
dimensions and generalizing it in the direction of extracting (quasi) convex, rather than merely quadratic elastic components, is feasible. These issues will be addressed systematically in a separate publication.
\vspace{1cm}
%%%%%%%%%%%%%%%%%%%%%%%%%%%%%%%%%%%%%%%%%%%%%%%%%%%%%%%%%%%%%%%%%%%

\appendix

\begin{flushleft}
{\large\bf Appendix}\\
\end{flushleft}

%\section{Appendix}
The following computations are largely based on the Eshelby's arguments presented in
\cite{ESHE49}. Consider a Volterra screw dislocation with zero-width core and with Burgers
vector $b$. The displacement $u_z(x,y)$ has the form
 \begin{equation}
 \label{eq:uzvolt}
 u_z(x,y)=\frac{b}{2\pi}\mathop{\rm Arg}(x+iy)=\frac{b}{2\pi}\arctan\frac{y}{x}+\frac{b}{2}\mathop{\rm
sign}(y)\theta(-x),
 \end{equation}
where $\theta$ is the Heaviside function, and where the indeterminacy in the
discontinuity of $u_z$ is resolved by specifying the glide plane ($y=0$).
The distributional part in the r.h.s.\ of Eq.\ (\ref{eq:uzvolt}),
usually omitted in the literature (e.g.\ \cite{HIRT82}), is crucial
to the present derivation because it represents the
irreversible atomic displacements on the plane $y=0$. We introduce
the eigendistortion, $\beta^*_{ij}$, as the part of
the dislocation-induced distortion $\beta_{ij}=u_{i,j}$, that is
\emph{not} linear-elastic. For our dislocation, its only non-zero
component is $\beta_{yz}^*(x,y)=b\,\theta(-x)\delta_D(y)$, where
$\delta_D$ is the Dirac distribution \cite{MURA87}.
The linear-elastic distortion, $\beta_{ij}^e$, is defined through the additive
decomposition of the total distortion $\beta_{ij}$, namely $\beta_{ij}^e\equiv
\beta_{ij}-\beta_{ij}^*$ \cite{MURA87}. The elastic strains are
$e_{ij}=\mathop{\rm sym}\beta_{ij}^e$. By using the identity
$[\arctan(1/x)]'=\pi\delta_D(x)-1/(1+x^2)$, we obtain that the distributional parts
in $\beta_{ij}$ and $\beta_{ij}^*$ mutually cancel out giving the standard result
\cite{HIRT82}
\begin{equation}
\label{eq:ezi} e_{xz}(x,y)=-\frac{b}{4\pi}\frac{y}{x^2+y^2},\quad
e_{yz}(x,y)=\frac{b}{4\pi}\frac{x}{x^2+y^2}.
\end{equation}
The stress induced by the eigenstrain is then
$\sigma_{iz}^*(x,y)=2\mu\, e_{iz}(x,y)$, where $i=x,y$.

In the presence of an applied shear stress \cite{NABA47}
$\sigma_a\equiv\sigma_{a\,yz}$, Eq.\ (\ref{eq:uzvolt}) becomes
\begin{equation}
\label{eq:uz2} u_z(x,y)=\frac{1}{\mu}\int_0^y {\rm
d}y'\,\sigma_a(x,y')+\frac{b}{2\pi}\arctan\frac{y}{x}+\frac{b}{2}\mathop{\rm
sign}(y)\theta(-x),
\end{equation}
The total stress is then $\sigma=\sigma^*+\sigma_a$ and $e=\sigma/2\mu$. Now, the key
step consists in averaging the stress over the layer of width $a$ containing the
glide plane. Introduce:
$
\overline{\sigma}_{ij}(x)\equiv \frac{1}{a}\int_{-a/2}^{+a/2} {\rm
d}y\, \sigma_{ij}(x,y).
$
From (\ref{eq:uz2}), the $x$ component of the total relative atomic lattice
displacement between the atomic planes at $y=\pm a/2$ reads:
\begin{equation}
\label{eq:duz} \delta(x)\!\equiv\! u_z(x,+a/2)\!-\!u_z(x,-a/2)\!
=\!\frac{a}{\mu}\!\left[\overline{\sigma}_a(x)\!+\!\frac{\mu b}{\pi
a}\arctan\left(\frac{a}{2x}\right)\right]\!+\!b\,\theta(-x).
\end{equation}
Furthermore on account of (\ref{eq:ezi}) the average shear stress
$\overline{\sigma}_{yz}(x)$ in the layer is:
\begin{equation}
\label{eq:ovs}
\overline{\sigma}_{yz}(x)=\overline{\sigma}_a(x)+\frac{\mu
b}{2\pi}\frac{1}{a}\int_{-a/2}^{a/2}{\rm
d}y\,\frac{x}{x^2+y^2}=\overline{\sigma}_a(x)+\frac{\mu b}{\pi
a}\arctan\left(\frac{a}{2x}\right).
\end{equation}
Comparison of (\ref{eq:ovs}) and(\ref{eq:duz}) shows that:
\begin{equation}
\label{eq:relue}
\delta(x)=(a/\mu)\overline{\sigma}_{yz}(x)+b\,\theta(-x).
\end{equation}

Consider next an Eshelby screw dislocation with an extended core described by a
continuous function $\eta(x)$. The distortion $\beta^*$ becomes:
$
\beta_{yz}^*(x,y)=\delta_D(y)\eta(x)=-\delta_D(y)\int_{x}^{+\infty}
{\rm d}x\,\eta'(x);
$
the Volterra dislocation corresponds to the limiting case
$\eta(x)=b\theta(-x)$. Displacements, strains and stresses are obtained by
convolution using ${\rm d}\beta_{yz}^*(x,y)\equiv -\delta_D(y)\eta'(x){\rm d}x$
\cite{ESHE49} as elementary distortions.  The analogs of Eqs.\
(\ref{eq:ovs},\ref{eq:relue}) are:
\begin{eqnarray}
\label{eq:ovs1}
\overline{\sigma}_{yz}(x)&=&\overline{\sigma}_a(x)-\frac{\mu}{\pi
a}\int_{-\infty}^{+\infty} {\rm
d}x'\,\arctan\left(\frac{a}{2(x-x')}\right)\,\eta'(x')\\
\label{eq:ovs2}
\delta(x)&=&(a/\mu)\overline{\sigma}_{yz}(x)+\eta(x).
\end{eqnarray}
Equation (\ref{eq:ovs2}), which we use in the paper, shows the
relation between the coarse-grained displacement $\delta$, and the
discontinuity $\eta$.
%%%%%%%%%%%%%%%%%%%%%%%%%%%%%%%%%%%%%%%%%%%%%%%%%%%%%%%%%%%%%%%%%%%

%%%%%%%%%%%%%%%%%%%%%%%% referenc.tex %%%%%%%%%%%%%%%%%%%%%%%%%%%%%%
% References
%%%%%%%%%%%%%%%%%%%%%%%% Springer-Verlag %%%%%%%%%%%%%%%%%%%%%%%%%%
% References must be sorted by alphabetical order
%%%%%%%%%%%%%%%%%%%%%%%%%%%%%%%%%%%%%%%%%%%%%%%%%%

%%%%%%%%%%%%%%%%%%%%%%%% referenc.tex %%%%%%%%%%%%%%%%%%%%%%%%%%%%%%
\end{document}